\documentstyle[12pt]{article}
\begin{document}
\begin{titlepage}
\begin{center} {\large \bf
ON A POSSIBLE EXPLANATION OF THE ORIGIN \\ OF THE QUARK MASS SPECTRUM}
\vglue 2cm
{\bf Dmitri Kazakov\footnote{E-mail: KazakovD@thsun1.jinr.dubna.su
\\ Supported in part by Russian Foundation for Fundamental Science, \\
Grant \# 94-02-03665-a }\\}
\vglue 1cm
{\it Bogoliubov Laboratory of
Theoretical Physics, Joint Institute for Nuclear Research, 141 980
Dubna, Moscow Region, RUSSIA \\}
\end{center}

\vspace{2cm}

\begin{abstract}
Contrary to the usual case when the quark mass spectrum is defined by
that of the quark Yukawa couplings while the vacuum expectation value
of the Higgs field remains universal, we suggest to consider the opposite
situation. Each generation of quarks and leptons is associated with its
own Higgs doublet, while the Yukawa couplings are the same. This situation
naturally arises in the framework of finite supersymmetric Grand
Unification scenario. Nondegeneracy of the quark mass spectrum is then
due to asymmetric vacuum expectation values which spontaneously  break
the discrete flavour symmetry in the Higgs sector.
\end{abstract}

\end{titlepage}

\section{Introduction}

The miracle of the quark and lepton mass spectrum has been a real
challenge for theorists for many years.  In the Standard Model the
masses appear as a result of spontaneous symmetry breaking and have the
form \begin{equation} m_i=y_i \cdot v \ \ \ ,  \label{1} \end{equation}
where $y_i$ are the corresponding Yukawa couplings to the Higgs field and
$v$ is the Higgs field vacuum expectation value.  Thus, the quark and
lepton mass spectrum is actually that of the Yukawa couplings.

How to get this spectrum and what is the reason of its nondegeneracy? The
answer to this question is still lacking. One should, however, have in
mind that the masses of quarks and leptons in eq.(\ref{1}) as well as all
the parameters in the Standard Model are the running ones. Therefore,
one usually talks about the value of the quark mass at the scale of the
mass itself, i.e. $m^2_q=\bar{m}^2_q(m^2_q)$. Thus we are interested
in the spectrum of the Yukawa couplings at the scale of quark masses.

Following the hypothesis of Grand Unification that the symmetry increases
at high energy it would be natural to assume that all Yukawa couplings are
equal at the unification scale and then split when coming to lower
energy scale thus defining the mass spectrum for quarks and leptons.
Indeed, this phenomenon really takes place for the masses of the
superpartners in the MSSM~\cite{1}.  However, the renormalization group
equations for the Yukawa couplings are different. Even if one assumes
that the flavour symmetry is slightly broken at high energy it will
 restore at lower energy scale, i.e.  the global flavour symmetry has
the property opposite to the Grand Unification of the local
symmetries~\cite{2}.  Therefore, to get the nondegenerate spectrum of
quark and lepton masses one has to input it at high energy.

One of the most interesting attempts of this kind is the one discussed in
ref.~\cite{3}, where the values of the Yukawa couplings as well as
the Kobayashi-Maskawa mixing matrix at the unification scale are given in
the form of the so-called {\em textures}, and then evolve to the observed
values at low energies. The textures themselves are chosen according to
maximal simplicity and symmetry while the needed parameters are fitted.
Without denying this possibility to get the quark mass spectrum, in
this paper we would like to suggest an alternative approach, which
naturally arises in attempts to construct SUSY GUTs free from ultraviolet
divergences~\cite{4},~\cite{5},~\cite{6}.

\section{SUSY GUT Scenario}

While the Standard Model exploits the minimal version of the Higgs
mechanism with only one Higgs doublet to provide masses to all quarks and
leptons simultaneously, already in the minimal supersymmetric extension of
the SM, the so-called MSSM, one needs at least two doublets. One doublet
then provides masses to up quarks, while the other - to down quarks and
leptons. Thus, we have two vacuum expectation values and their ratio
$\tan\beta \equiv v_2/v_1$ is the free parameter of the model.

In the standard minimal SUSY GUT scenario~\cite{1} the theory possesses
both the supersymmetry and the unified gauge symmetry at the
unification scale with soft SUSY breaking terms arising from a
supergravity. At this scale all quarks and leptons are massless and
their superpartners all have the same mass. Going down to lower
energies the superpartners masses run according to the RG equations,
split due to different interactions and, thus, give us the mass
spectrum at Tev scale. This is accompanied also by the radiative
spontaneous symmetry breaking, which leads to the reconstruction of the
vacuum state. The latter, according to the usual Higgs mechanism,
provides us with the masses for quarks, leptons and $SU(2)$ gauge
bosons and additional mass terms to their superpartners.

Quarks and leptons themselves are not involved in this process, since they
are relatively light. Their mass spectrum remains completely arbitrary due
to the arbitrariness of the corresponding Yukawa couplings. Having two
Higgs vacuum expectation values with arbitrary $\tan\beta $ fitted by
experiment does not change the situation. However, already here the value
of $\tan\beta $ can be found from the minimization of the potential for
neutral Higgses, if the parameters are known, and differs from
unity~\cite{1}.  Thus, we can get a hierarchy if the potential has
various minima, though it is not essential  when the Yukawa couplings
remain arbitrary.

\section{Finite SUSY Models}

Whence we have already enlarged the number of Higgses, we can go
further and consider some non-minimal model. At first sight this looks
absolutely hopeless because of increasing number of arbitrary
parameters. However, there is one exception. This is a SUSY GUT model
which though non-minimal still remains almost as rigid as the
minimal one. It is distinguished by its ultraviolet properties being
absolutely UV finite to all orders of perturbation
theory~\cite{4},~\cite{5}.  Let us remind the main properties of a
finite SUSY GUT:
\begin{itemize}
\item the number of generations is fixed by the requirement of finiteness,
\item the representations and the number of the Higgs fields are fixed,
\item all the Yukawa couplings are expressed in terms of the gauge one,
\item the various realistic possibilities are  given by
$SU(5),SU(6), SO(10)$ and $E(6)$ gauge groups with few generations.
\end{itemize}

We consider below the simplest case of the gauge group
$SU(5)$~\cite{6}.  Then only three generations are allowed and the
Higgs sector contains one $24$ representation, which breaks $SU(5)$
down to $SU(3)\times SU(2) \times U(1)$, and four pairs of Higgses in
$5$ and $\bar{5}$ representations. Thus we have three extra pairs of
Higgses compared to the MSSM.

After spontaneous breaking of $SU(5)$ one naturally achieves that one
pair of Higgses obtains the mass of the order of $M_X$, while the other
three split into doublets and triplets under $SU(2)$. Triplets become
heavy while the doublets remain light due to the fine tuning. As a
result below $M_X$ we get three pairs of light Higgs doublets: one for
each generation.

Equation (\ref{1}) for the quark masses is now modified. As we have
already mentioned, all the Yukawa couplings are uniquely defined by the
requirement of finiteness at the GUT scale. In the leading order of PT
one has:
\begin{equation} y_{D_i}= const\cdot g, \ \ \ y_{U_i}=
\sqrt{\frac{4}{3}}\  const \cdot g, \ \ \ y_{L_i}= const \cdot g \ ,
\label{2} \end{equation}
where $y_{D_i},y_{U_i}$ and $y_{L_i}$ are the
Yukawa couplings of down and up quarks and leptons, respectively, and
g is the $SU(5)$ gauge coupling.  In higher orders one has the
calculable corrections to eq.(\ref{2}) as a power series over $g$.

These values of the Yukawa couplings
serve as the boundary conditions for the RG equations. Since the
interactions are flavour symmetrical, the values of the Yukawa couplings
at $M_Z$ are also flavour degenerate.

Then eq.(\ref{1})  takes the form
\begin{equation}
m_{U_i}=y_U \cdot v_i,\ \ \ m_{D_i}=y_D \cdot \bar{v}_i,\ \ \
 m_{L_i}=y_L \cdot \bar{v}_i\ , \label{3}
\end{equation}
where $v_i$ and $\bar{v}_i$ ($i=1,2,3$) are the v.e.v.s of the Higgs
fields in $2$ and $\bar{2}$ representations, respectively.

\section{Quark Mass Spectrum}

As one can see from eq.(\ref{3}) the mass spectrum of quarks and
leptons is now defined by the v.e.v.s rather than by the Yukawa
couplings. In its turn the v.e.v.s themselves are the solutions of the
minimization conditions for the Higgs potential.

At the GUT scale we start with the potential
\begin{equation}
V(H_i,\bar{H}_i)=V_{SUSY} \ + \ V_{Soft} \ , \label{4}
\end{equation}
which has a discrete symmetry of interchange $H\leftrightarrow\bar{H}$
and the generation symmetry. However, when running the parameters to
the lower energies where spontaneous breaking of $SU(2)$ gauge
invariance takes place, both these symmetries are destroyed.  The
reasons for this are two fold: different renormalization of $H$ and
$\bar{H}$ fields and the Higgs mixing matrix $\mu_{ij}$, which is
analogous to the Kobayashi-Maskawa mixing matrix of quarks but in the
Higgs sector.

At the electroweek scale the potential for the neutral Higgs components
takes the form:
\begin{eqnarray}
V &=& m_1^2|H_i|^2+m_2^2|\bar{H}_i|^2-2\bar{H}_i\mu_{ij}H_j \label{5}
\\ &+& \frac{g^2+{g'}^2}{8}\left(|\bar{H}_i|^2-|H_i|^2\right)^2, \ \ \
\ \ (i=1,2,3) .\nonumber
\end{eqnarray}
Looking for the minima of the
potential (\ref{5}) one can find several solutions which are degenerate
and spontaneously break the generation symmetry. Since this symmetry
due to the presence of the mixing matrix $\mu_{ij}$ becomes
discrete, no Goldstone bosons appear.  Each solution creates the
hierarchy of v.e.v.s and, hence, the hierarchy of masses.

This phenomenon happens even if $\mu_{ij}$ is diagonal, but
asymmetry of $\mu_{ij}$ is needed to avoid the global $SO(3)$
generation invariance and appearance of Goldstone bosons.

To illustrate the idea we consider the simplified example with two
Higgses, just like in the MSSM~\cite{1}. One has:
\begin{equation}
V=m_1^2H^2+m_2^2\bar{H}^2-2\mu \bar{H}H+\frac{g^2}{8}(\bar{H}^2-H^2)^2 .
\label{6}  \end{equation}
The minimization conditions are
\begin{eqnarray}
\frac{\delta V}{\delta H}&=& 2m_1^2H-2\mu \bar{H}-\frac{g^2}{2}(\bar{H}^2
-H^2)H = 0 , \nonumber   \\
 & & \label{7} \\
\frac{\delta V}{\delta \bar{H}}&=& 2m_2^2\bar{H}-2\mu
H+\frac{g^2}{2}(\bar{H}^2 -H^2)\bar{H} = 0  \nonumber
\end{eqnarray}
Introducing the vacuum expectation values
\begin{eqnarray}
<H> &=& v_1 = v \cos\beta , \nonumber \\
<\bar{H}> &=& v_2 = v \sin\beta , \nonumber
\end{eqnarray}
the solution to eqs.(\ref{7}) has the form
\begin{equation}
v^2=\frac{4}{g^2}\frac{m_1^2-m_2^2\tan^2\beta }{\tan^2\beta -1}, \ \ \
\sin 2\beta = \frac{2\mu }{m_1^2+m_2^2} ,  \label{8}
\end{equation}
or
\begin{equation}
\tan\beta \equiv \frac{v_2}{v_1}=
\frac{m_1^2+m_2^2\pm\sqrt{(m_1^2+m_2^2)^2-4\mu^2}}{2\mu }. \label{9}
\end{equation}
The sign in eq.(\ref{9}) depends on relative values of $m_1^2$ and
$m_2^2$. One takes $(+)$ if $m_1^2> m_2^2$ and $(-)$ in the opposite
case. One of the solitions being the inverse of the other. When
$m_1^2=m_2^2$ only the trivial solution, $v_1=v_2=0$, exists. Evolving
the difference between $m_1^2$ and $m_2^2$ we
spontaneously break the discrete flavour symmetry in the Higgs sector
and create the hierarchy and, hence, the mass spectrum.  Therefore,
even in a symmetrical original potential at the GUT scale one can have
asymmetric solutions.

The Higgs particles also obtain masses which are given by the
diagonalization of the matrix of the second derivatives of the
potential (\ref{6}) and have the form:
\begin{eqnarray}
m^2_{H_{1,2}}&=&\frac{1}{2}\left[m_1^2+m_2^2+M_Z^2 \right.\nonumber \\
&& \left.\pm
\sqrt{(m_1^2+m_2^2+M_Z^2)^2-4(m_1^2+m_2^2)M_Z^2\cos^22\beta }\right],
\label{10} \end{eqnarray}
where $M_Z^2=g^2v^2/2$. One of the Higgses
can be light while the other is heavy.

The same mechanism works in  a realistic model. All the parameters
defining the spectrum of masses, like $m_1^2,m_2^2,\mu _{ij}$, etc are
then determined from the requirement of consistency of the model as
in the MSSM ~\cite{1}. The lightest Higgs particle plays the role of a
single Higgs of the SM.

\section{Conclusion}

We have demonstrated that it is possibile to obtain a quark mass
spectrum which arises as a result of spontaneous breaking of $SU(2)$
symmetry by different v.e.v.s of the Higgs fields. The generation
symmetry is reduced to the discrete one by the mixing in the Higgs
sector and then is spontaneously broken. One does not need to introduce
the quark spectrum "by hand" either at low energy, or at the GUT
scale. Since all the Yukawa couplings in the finite model are known at
the unification scale and then run according to the known RG
equations, the only free parameters coming from the Higgs
potential, the mixing matrix $\mu_{ij}$ and the soft SUSY breaking
terms. The number of arbitrary parameters does not exceed that of
the MSSM and can be even less if the mixing in the Higgs sector is
correlated (or identified) with that in the quark sector.

The detailed analysis within the SUSY GUT scenario is in progress and
will be published elsewhere.

\end{document}